\documentclass[twocolumn,aps,prl]{revtex4-1}
\usepackage{bm}
\usepackage{amsfonts}
\usepackage{amssymb}
\usepackage{amsmath}
\usepackage{graphicx}

\begin{document}

\title{Uncertainty relation for photons}
\author{Iwo Bialynicki-Birula}\email{birula@cft.edu.pl}
\affiliation{Center for Theoretical Physics, Polish Academy of Sciences\\
Al. Lotnik\'ow 32/46, 02-668 Warsaw, Poland}
\author{Zofia Bialynicka-Birula}\affiliation{Institute of Physics, Polish Academy of Sciences\\
Al. Lotnik\'ow 32/46, 02-668 Warsaw, Poland}

\begin{abstract}
The uncertainty relation for the photons in three dimensions that overcomes the difficulties caused by the nonexistence of the photon position operator is derived in quantum electrodynamics. The photon energy density plays the role of the probability density in configuration space. It is shown that the measure of the spatial extension based on the energy distribution in space leads to an inequality that is a natural counterpart of the standard Heisenberg relation. The equation satisfied by the photon wave function in momentum space which saturates the uncertainty relations has the form of the Schr\"odinger equation in coordinate space in the presence of electric and magnetic charges.
\end{abstract}

\maketitle

The uncertainty relations that first appeared in the fundamental Heisenberg paper on the conceptual content of quantum mechanics \cite{heis} play an important role in elucidating specific properties of the quantum world. The aim of this Letter is to derive in the framework of quantum electrodynamics a sharp uncertainty relation for photons treated as three-dimensional objects. The photon wave function that saturates our inequality has already appeared in different contexts.

An uncertainty relation for the momentum and the center of mass coordinate along {\em a given direction} can easily be obtained from the commutators between the generators of the Poincar\'e group. It follows from general principles \cite{dirac0} that the commutators between the components of the momentum and the Lorentz boost are:
\begin{align}\label{dir}
[{\hat N}_i,{\hat P}_j]=i\hbar\delta_{ij}{\hat H}.
\end{align}
Therefore, each component of the center of mass coordinate (the boost operator divided by the energy operator with proper symmetrization to secure Hermiticity),
\begin{align}\label{r}
{\hat R}_i=\textstyle\frac{1}{2}\left({\hat H}^{-1}{\hat N}_i+{\hat N}_i{\hat H}^{-1}\right),
\end{align}
is a canonically conjugate variable to the corresponding component of momentum, $[{\hat R}_i,{\hat P}_j]=i\hbar\delta_{ij}$. Hence, the standard Heisenberg relation $\Delta x_i^2\Delta p_i^2\ge\hbar^2/4$ follows separately for every direction, with Gaussian functions saturating each inequality. These simple one-dimensional uncertainty relations hold for {\em every relativistic system}. In the special case of photons they were obtained by Holevo \cite{h} in the framework of estimation theory. Holevo also proved that ``the transversality condition precludes the three inequalities from becoming equalities simultaneously.'' This was to be expected because the existence of one wave function that saturates all three standard Heisenberg uncertainty relations would mean that photons are not any different from nonrelativistic particles. In a general approach based on the properties of the Poincar\'e group, the impossibility to saturate simultaneously the three uncertainty relations follows from the noncommutativity of the components of ${\bm{\hat R}}$,
\begin{align}\label{com}
[{\hat R}_i,{\hat R}_j]=-i\hbar c^2{\hat H}^{-1}\left({\hat M}_{ij}-{\hat R}_i{\hat P}_j+{\hat R}_j{\hat P}_i\right){\hat H}^{-1}.
\end{align}
The expression in parentheses on the right side represents the spin: the total angular momentum minus the orbital part. Schwinger \cite{js} has shown that for massless particles with helicity $\lambda$ the right-hand side in (\ref{com}) becomes:
\begin{align}\label{js}
{\bm{\hat R}}\times{\bm{\hat R}}=-i\lambda{\bm k}/k^3.
\end{align}
He also found a very rough estimate of the lower bound in the three-dimensional uncertainty relation.

In what follows we fully realize the Schwinger program by deriving a precise form of the uncertainty relation for photons in three dimensions. We also find the photon wave function in momentum space that saturates the inequality and we confirm the mathematical analogy with magnetic monopoles discovered by Schwinger. Our uncertainty relation for photons has the form:
\begin{align}\label{one}
\Delta r\Delta p\ge 4\hbar,
\end{align}
and is saturated (i.e. becomes an equality) by---what might be called---the fundamental solution of Maxwell equations. The lower bound in this relation is 4 times larger than the Schwinger estimate. It is also larger than $3/2\hbar$ that would be obtained by blindly extending the one-dimensional relation for photons to three dimensions. The larger value is due, of course, to the impossibility to saturate simultaneously the one-dimensional uncertainty relations for photons in all three directions.

In order to overcome the problem of the nonexistence of a bona fide photon position operator and the associated probability density in coordinate space we rely on the photon {\em energy density}. To measure the photon spread in coordinate space we use the second moment of the photon energy distribution---the generator $K_0$ of a conformal transformation \cite{bb}.

There is no problem with the momentum distribution of photons since there exists a well defined photon wave function in momentum space. This function appears in the representation theory of the Poincar\'e group expounded by Wigner \cite{wig}. In order to describe completely the state of a photon we need two such functions, say $f_\pm(\bm k)$, that refer to two helicities (two circular polarizations). Under all proper Poincar\'e transformations these two functions transform separately but they are interchanged under reflections. In quantum theory these functions (normalized to 1) become the probability amplitudes for finding the photon with momentum $\bm k$ and the helicity $\lambda=+1$ or $\lambda=-1$. Following Wigner, we choose the relativistic form of the norm,
\begin{align}\label{norm}
||f||^2=\sum_{\lambda}\int\!\frac{d^3k}{k}\,|f_\lambda(\bm k)|^2,
\end{align}
where we allow for both polarizations to be present.

The straightforward Fourier transformation of the photon wave function in momentum space does not produce a local function in coordinate space \cite{lp,p}. To describe the localization properties in coordinate space we shall rely solely on the {\em electromagnetic fields}.

It is convenient to express the energy density in terms of a complex combination of the electric and magnetic field operators, named in \cite{pwf} the Riemann-Silberstein vector,
\begin{align}\label{F}
{\hat{\bm F}}(\bm r,t)=\frac{{\hat{\bm D}}(\bm r,t)}{\sqrt{2\epsilon}}+i\frac{{\hat{\bm B}}(\bm r,t)}{\sqrt{2\mu}}.
\end{align}
The energy density operator is equal to ${\hat{\bm F}}^\dagger(\bm r,t)\!\cdot\!{\hat{\bm F}}(\bm r,t)$.

We shall take the following expectation value:
\begin{align}\label{r2}
\Delta r=\frac{1}{\hbar c}\langle\int\!d^3r\,{\bm r}^2\,:{\hat{\bm F}^\dagger}(\bm r,t)\!\cdot\!{\hat{\bm F}}(\bm r,t):\rangle
\end{align}
as a measure of the spread in the coordinate space. In this way we follow closely the concept of photon localization underlying the theory of photodetection \cite{rg,mw}. According to this theory, the localization of photons is determined by the correlation functions of the  field operators. Normal ordering in (\ref{r2}) eliminates the contribution from the vacuum state. We have chosen the origin of the coordinate system at the center of mass ($\langle\bm r\rangle=0$). Despite the presence of ${\bm r}^2$, $\Delta r$ has the dimension of length like its counterpart in the Heisenberg uncertainty relation.

In quantum electrodynamics ${\hat{\bm F}}(\bm r,t)$ has the following representation in terms of the annihilation and creation operators for photons with positive and negative helicities $a_+(\bm k)$ and $a_-^\dagger(\bm k)$ (cf. for example, \cite{pwf,bb}),
\begin{align}\label{rep}
&{\hat{\bm F}}(\bm r,t)=\sqrt{\hbar c}\int\!\frac{d^3k}{(2\pi)^{3/2}}\nonumber\\
&\times{\bm e}(\bm k)\left[a_+(\bm k)e^{i\bm k\cdot\bm r-i\omega t}+a_-^\dagger(\bm k)e^{-i\bm k\cdot\bm r+i\omega t}\right].
\end{align}
The normalized polarization vector ${\bm e}(\bm k)$ is taken as:
\begin{eqnarray}\label{polar}
{\bm e}({\bm k}) = \frac{1}{\sqrt{2}\,k \sqrt{k_x^2+k_y^2}}\left(
\begin{array}{c}
-k_x k_z+i k k_y\\
-k_y k_z-i k k_x\\
k_x^2+k_y^2
\end{array}
\right).
\end{eqnarray}
The polarization vector ${\bm e}(\bm k)$ always distinguishes a certain direction. We have chosen here this direction along the $z$ axis. The commutation relations between the creation and annihilation operators,
\begin{align}\label{cr}
\left[a_\lambda(\bm k),a_{\lambda'}^\dagger(\bm k')\right]=\delta_{\lambda\lambda'}k\,\delta^{(3)}(\bm k-\bm k'),
\end{align}
follow from the commutation relations for field operators.

We shall evaluate $\Delta r$ in the one-photon state that is created from the vacuum state by the following combination of creation operators:
\begin{align}\label{1p}
|f\rangle=\int\!\frac{d^3k}{k}\left[f_+(\bm k)a_+^\dagger(\bm k)+f_-(\bm k)a_-^\dagger(\bm k)\right]|0\rangle.
\end{align}
The norm of this state vector is given by Eq.~(\ref{norm}). We will not assume now that the wave functions are normalized to 1 since they will serve later as variational variables. Our measure of the uncertainty in position for this state reduces to the second moment of the classical energy density (divided by $\hbar c$),
\begin{align}\label{r2c}
\Delta r =\frac{1}{\hbar c||f||^2}\int\!d^3r\,{\bm r}^2{\bm F}^*(\bm r,t)\!\cdot\!{\bm F}(\bm r,t),
\end{align}
where the c-number Riemann-Silberstein vector ${\bm F}(\bm r,t)$ is built from the photon wave functions $f_\pm(\bm k)$,
\begin{align}\label{rep1}
&{\bm F}(\bm r,t)=\sqrt{\hbar c}\int\!\frac{d^3k}{(2\pi)^{3/2}}\nonumber\\
&\times{\bm e}(\bm k)\left[f_+(\bm k)e^{i\bm k\cdot\bm r-i\omega t}+f_-^*(\bm k)e^{-i\bm k\cdot\bm r+i\omega t}\right].
\end{align}
We may choose $t=0$, since there is no preferred origin of time. Then, $\Delta r$ can be converted into an integral in momentum space,
\begin{align}\label{intk}
\Delta r=\frac{1}{||f||^2}\sum_\lambda\int\!d^3k \left[|{\bm D}_\lambda f_\lambda|^2+k^{-2}|f_\lambda|^2\right],
\end{align}
where $\bm D_\lambda$ is the covariant derivative (in momentum space) on the light cone \cite{as,bb,bb1}. The helicity $\lambda$ plays here the role of the charge,
\begin{align}\label{cder}
\bm D_\lambda&={\bm\nabla}-i\lambda{\bm\alpha}({\bm k}),\\
{\bm\alpha}({\bm k})&=\frac{k_z}{k\,(k_x^2+k_y^2)}\left(-k_y,k_x,0\right).\label{alpha}
\end{align}
In the derivation of (\ref{intk}) we used the formula:
\begin{align}\label{form}
&x_i{F}_j(\bm r,0)=i\sqrt{\hbar c}\int\!\frac{d^3k}{(2\pi)^{3/2}}\nonumber\\
&\!\!\times\Big[e^{i\bm k\cdot\bm r}\partial_i\left(e_j(\bm k)f_+(\bm k)\right)-e^{-i\bm k\cdot\bm r}\partial_i\left(e_j(\bm k)f_-^*(\bm k)\right)\Big],
\end{align}
and the following properties of the polarization vector:
\begin{align}\label{prop}
i{\bm e}^*(\bm k)\!\cdot\!\partial_i{\bm e}(\bm k)&=\alpha_i(\bm k),\quad\quad\;{\bm e}(\bm k)\!\cdot\!\partial_i{\bm e}(\bm k)=0,\\
\partial_i{\bm e}^*(\bm k)\!\cdot\!\partial_i{\bm e}(\bm k)&=\frac{1}{k_x^2+k_y^2},\;\;
\partial_i{\bm e}(\bm k)\!\cdot\!\partial_i{\bm e}(\bm k)=0.
\end{align}

The vector $\alpha(\bm k)$ plays the role of the connection on the light cone that defines the parallel transport for the photon wave function in momentum space. When the phase of $f_\lambda(\bm k)$ is changed by $\exp(i\lambda\phi(\bm k))$, the phase of ${\bm e}(\bm k)$ must undergo a compensating change, ${\bm e}(\bm k)\to \exp[-i\phi(\bm k)]{\bm e}(\bm k)$, to keep the field ${\bm F}(\bm r,t)$ intact. This results in a gauge transformation ${\bm\alpha}(\bm k)\to{\bm\alpha}(\bm k)+\nabla\phi(\bm k)$. Note, that the covariant derivative vector $i\bm D_\lambda$ obeys the same commutation relations as the operator ${\bm{\hat R}}$,
\begin{align}\label{dcr}
i{\bm D}_\lambda\times i{\bm D}_\lambda=i\lambda\nabla\times{\bm\alpha}(\bm k)=-i\lambda\bm k/k^3.
\end{align}
Therefore it can be viewed as a representation of ${\bm{\hat R}}$ in momentum space.

The last equation means that ${\bm\alpha}(\bm k)$ can be treated as a vector potential of a magnetic monopole. Such a vector is always singular on the Dirac string \cite{dirac}. Our choice of the polarization vector ${\bm e}(\bm k)$ leads to the string along the whole $z$ axis \{cf. for example, Eq.~(16) of Ref.~\cite{mklg}\}.

Our uncertainty relation for photons involves the product $\Delta r\Delta p$ like the standard Heisenberg relation. The natural choice for $\Delta p$ is:
\begin{align}\label{pe}
\Delta p =\frac{1}{||f||^2}\sum_{\lambda}\int\!\frac{d^3k}{k}\,\hbar \,|{\bm k}-\langle{\bm k}\rangle|\;|f_\lambda(\bm k)|^2.
\end{align}
To simplify the calculations we shall choose the coordinate system comoving with the wave packet in such a way that $\hbar\langle{\bm k}\rangle=0$ (center of momentum frame). Then,
\begin{align}\label{pe1}
\Delta p =\frac{\hbar}{||f||^2}\sum_{\lambda}\int\!\frac{d^3k}{k}\,k|f_\lambda(\bm k)|^2=\hbar\langle k\rangle.
\end{align}
The existence of a finite lowest value of the product $\Delta r\Delta p$ is an expression of the uncertainty relation for the photon. This value will be calculated by a variational method.

We test this method first for the standard Heisenberg uncertainty relation. Again, we choose the center of mass and the center of momentum frame, so that $\langle{\bm r}\rangle=0$ and $\langle{\bm p}\rangle=0$. The expression to be minimized is:
\begin{align}\label{heis}
\varpi^2=\frac{\sigma_r^2\sigma_p^2}{\hbar^2}=\frac{\int\!d^3r\,{\bm r}^2\psi^*({\bm r})\psi({\bm r})}{\int\!d^3r\,\psi^*({\bm r})\psi({\bm r})}
\frac{\int\!d^3r\,\nabla\psi^*({\bm r})\!\cdot\!\nabla\psi({\bm r})}
{\int\!d^3r\,\psi^*(x)\psi(x)}.
\end{align}
Variation with respect to $\psi^*({\bm r})$ leads to the Schr\"odinger equation for the three-dimensional harmonic oscillator,
\begin{align}\label{ho}
{\textstyle\frac{1}{2}}\left[-\sigma_{\bm r}^2\Delta+\varpi^2\,{\bm r}^2/\sigma_{\bm r}^2\right]\psi({\bm r})=\varpi^2\,\psi({\bm r}).
\end{align}
where we assumed at the end that the wave function is normalized to 1. The operator appearing on the left-hand side has the eigenvalues $\varpi(n+3/2)$. Comparing this with the right-hand side, we obtain $\varpi=(n+3/2)$. The lowest value of $\varpi$ is $3/2$ and it leads to the standard Heisenberg uncertainty relation in three dimensions,
\begin{align}\label{ur0}
\sigma_r\sigma_p\ge \frac{3}{2}\,\hbar.
\end{align}
The wave function of the ground state---the Gaussian---saturates this inequality. Owing to the linearity of the variational equation (\ref{ho}) we did not have to use a Lagrange multiplier. Note that for the wave functions that saturate the inequality the uncertainties in position and momentum are evenly distributed, when measured in conjugate units, namely $\sigma_r=\sqrt{3/2}\,a,\;\sigma_p=\sqrt{3/2}\,\hbar/a$.

Now, we apply an analogous variational procedure to the dimensionless expression $\gamma=\Delta r\Delta p/\hbar$. With the use of (\ref{cder}) and (\ref{alpha}) we rewrite the formulas (\ref{intk}) and (\ref{pe1}) in spherical coordinates,
\begin{align}
&\Delta r =\frac{1}{||f||^2}\sum_\lambda\!\!\int_0^\infty\!\!\!\!\!dk k^2\!\int_0^\pi\!\!\!\!
d\theta\sin\theta\!\! \int_0^{2\pi}\!\!\!\!\!d\varphi \Big[|\partial_kf_\lambda|^2+\frac{|\partial_\theta f_\lambda|^2}{k^2}\label{sphericalr}\nonumber\\
&+\frac{|\partial_\varphi f_\lambda|^2+|f_\lambda|^2+i\lambda\cos\theta\left(f^*_\lambda\partial_\varphi f_\lambda-f_\lambda\partial_\varphi f^*_\lambda\right)}{k^2\sin^2\theta}\Big],\\
&\Delta p =\frac{\hbar}{||f||^2}\sum_\lambda\!\!\int_0^\infty\!\!\!\!\!dkk^2\!\int_0^\pi\!\!\!\!
d\theta\sin\theta\!\! \int_0^{2\pi}\!\!\!\!\!d\varphi\,|f_\lambda|^2.\label{sphericalk}
\end{align}
Variation of $\gamma$ with respect to $f^*_\lambda(k,\theta,\varphi)$ produces the following equation:
\begin{widetext}
\begin{align}\label{eq}
\left[-\frac{1}{\kappa^2}\partial_\kappa \kappa^2\partial_\kappa-\frac{1}
{\kappa^2\sin\theta}\partial_\theta\sin\theta\,\partial_\theta
+\frac{1}{\kappa^2\sin^2\theta}
\left(-\partial_\varphi^2+1+2i\lambda\cos\theta\,\partial_\varphi
\right)-\frac{2\gamma}{\kappa}+\gamma\right]f_\lambda(\kappa,\theta,\varphi)=0,
\end{align}
\end{widetext}
where we assumed (after performing the variation) that the norm of $f$ is equal to 1 and we introduced a dimensionless variable $\kappa=\hbar k/\Delta p$. After the separation of variables, $f_\lambda(\kappa,\theta,\varphi)={\mathcal K}(\kappa)\Theta_\lambda(\theta)\exp(im\varphi)$, we obtain:
\begin{align}\label{eqk}
\left[-\frac{1}{\kappa^2}\partial_\kappa \kappa^2\partial_\kappa+\frac{j(j+1)}{\kappa^2}-\frac{2\gamma}{\kappa}\right]{\mathcal K}(\kappa)=-\gamma{\mathcal K}(\kappa),
\end{align}
\begin{align}\label{eqt}
\left[-\frac{1}{\sin\theta}\,\partial_\theta\sin\theta\,\partial_\theta
+\frac{m^2+\lambda^2-2\lambda m\cos\theta}{\sin^2\theta}\right]
\Theta_\lambda(\theta)\nonumber\\=j(j+1)\Theta_\lambda(\theta).
\end{align}

The equation for $\Theta_\lambda(\theta)$ is exactly the same as in the theory of magnetic monopoles (cf. for example, \cite{mklg,milton}). We wrote $\lambda^2$ instead of 1 to have the form of Eq.~(\ref{eqt}) identical with the monopole case. In this equation $\lambda$ plays the role of the product of the electric and magnetic charges divided by $\hbar c$. The presence of the monopole string along the $z$ axis excludes the value $j=0$. The contributions to $\Delta r$ and to $\Delta p$ from the wave functions with $\lambda=\pm 1$ are equal and independent. For definiteness, we choose $\lambda=1$.

The equation for the radial part has the same form as the Schr\"odinger equation for the Coulomb problem. Hence, the spectrum of the operator appearing on the left-hand side of Eq.~(\ref{eqk}) is $-\gamma^2/(n_r+j+1)^2$. Since this must be equal to $-\gamma$, we obtain $\gamma=(n_r+j+1)^2$. We are interested in the smallest value of $\gamma$ which corresponds to $n_r=0,\;j=1$. This gives $\gamma=4$ as in Eq.~(\ref{one}).

There are three independent solutions of Eq.~(\ref{eqt}) for $j=1$ that saturate the uncertainty relation. In the angular momentum basis they correspond to $m=0,\,\pm 1$ but we found the Cartesian basis more convenient. Normalized solutions of the variational equation that saturate the uncertainty relation are:
\begin{subequations}\label{gs}
\begin{align}
f^{(z)}(k,\theta,\varphi)&=\frac{a^2}{\sqrt{\pi}}k\sin\theta\,e^{-k a},\\
f^{(x)}(k,\theta,\varphi)&=\frac{a^2}{\sqrt{\pi}}
k(-i\sin\varphi-\cos\theta\cos\varphi)\,e^{-k a},\\
f^{(y)}(k,\theta,\varphi)&=\frac{a^2}{\sqrt{\pi}}
k(i\cos\varphi-\cos\theta\sin\varphi)\,e^{-k a},
\end{align}
\end{subequations}
where $a$ is an arbitrary scaling parameter. These three solutions are not essentially different. They correspond to three choices of the polarization vector ${\bm e}(\bm k)$ obtained from (\ref{polar}) by cyclic permutations of the coordinate axes: $k_z\to k_x \to k_y \to k_z$.

The uncertainty relation (\ref{one}) does not depend on $a$. Moreover, the uncertainties in position and momentum for the wave functions that saturate the inequality are again, as is the case for the Heisenberg relation, equally distributed: $\Delta_r=2\,a,\;\Delta_p=2\,\hbar/a$. The larger value of the numerical coefficient in the uncertainty relation for photons than in the Heisenberg relation (\ref{ur0}) is due to the magnetic monopole term in Eq.~(\ref{eqt}). Since this term excludes $s$ waves, the maxima of the functions (\ref{gs}) are shifted away from zero.

The solution of Maxwell equations that saturates the uncertainty relation deserves to be named the fundamental solution. It can be built from the second derivatives (Whittaker construction \cite{whit}) of the complexified fundamental solution $\Phi$ of the d'Alembert equation,
\begin{align}\label{fs}
\Phi(\bm r,\tau)=\left[{\bm r}^2-(\tau-ia)^2\right]^{-1},
\end{align}
where $\tau=ct$. For our choice of ${\bm e}(\bm k)$, the wave function $f^{(z)}$ leads to:
\begin{align}\label{whit}
{\bm F}^{(z)}(\bm r,\tau)=\frac{a^2\sqrt{\hbar c}}{4\pi^2}\left(
\begin{array}{c}\partial_x\partial_z+i\partial_y\partial_\tau\\
\partial_y\partial_z-i\partial_x\partial_\tau\\
-\partial_x^2-\partial_y^2\end{array}\right)\Phi(\bm r,\tau).
\end{align}
The superscript $(z)$ indicates that the $z$ axis was singled out in the definition of the polarization vector. The remaining two solutions, built from $f^{(x)}$ and $f^{(y)}$, are obtained from (\ref{whit}) by cyclic permutations of partial derivatives. We may also apply dual rotations to further enlarge the set of solutions of Maxwell equations that saturate the uncertainty relation. This whole set may be viewed as just one fundamental solution acted upon with various symmetry transformations. At the moment of maximal compression ($t=0$) they all have the same spherically symmetric energy density ${\cal E}$:
\begin{align}\label{enp}
{\cal E}=\hbar c\frac{16a^4}{\pi^2}\frac{1}{(a^2+r^2)^4}.
\end{align}
The simplest solution, given by the expression (\ref{whit}), reduces at $t=0$ to a pure electric field:
\begin{align}\label{el}
{\bm E}=\sqrt{\hbar c}\frac{4a^2}{\pi}\frac{\left(2zx-2ay,2yz+2ax,a^2-r^2+2z^2\right)}{(a^2+r^2)^4},
\end{align}
whose all field lines are ellipses.

The fundamental solution of Maxwell equations appeared in various investigations. Its exceptional properties were recognized by Synge \cite{synge} who called it ``an electromagnetic model of a material particle.'' Later, this solution appeared as the main ingredient in the construction of electromagnetic knots by Hopf fibrations \cite{at,r,rt,ib} and in the study of vortex lines embedded in the solutions of Maxwell equations \cite{ibb}. It was even used to explain the ball lightning \cite{rt1}.

\acknowledgments

We thank the referee for informing us about the Holevo papers. This research was partly supported by the grant from the Polish Ministry of Science and Higher Education for the years 2010--2012.

\end{document}